\documentclass[showpacs,amsmath,amssymb, aps]{revtex4}

\usepackage{tabularx}
\usepackage{graphicx}
\usepackage{float}
\usepackage{dcolumn}
\usepackage{bm}
\usepackage{mathrsfs}
\usepackage{amsmath}

\begin{document}


\title{Fuzzy dark matter black holes and droplets}
\author{D. Batic}
\email{davide.batic@ku.ac.ae}
\affiliation{%
Department of Mathematics,\\  Khalifa University of Science and Technology,\\ Main Campus, Abu Dhabi,\\ United Arab Emirates}
\author{D. Asem Abuhejleh}
\email{100045070@ku.ac.ae}
\affiliation{%
Department of Mathematics,\\  Khalifa University of Science and Technology,\\ Main Campus, Abu Dhabi,\\ United Arab Emirates}
\author{M. Nowakowski}
\email{mnowakos@uniandes.edu.co}
\affiliation{
Departamento de Fisica,\\ Universidad de los Andes, Cra.1E
No.18A-10, Bogota, Colombia
}

\date{\today}

\date{\today}

\begin{abstract}
{\bf{We consider the possibility of having Dark Matter (DM) black holes motivated by the Einasto density profile.  This generalizes both the noncommutative mini black hole model and allows DM to enter as the matter constituent which makes up the black hole. We show that it is possible to construct a black hole solution for each value of the Einasto index and for different values of the mass parameter, provided that the we work with the energy-momentum tensor of an anisotropic fluid. In particular, we achieve that by first considering the equation of state (EOS) $p_r=-\rho$. It turns out that the corresponding black hole  solution exhibits a horizon structure similar to that of a Reissner-Nordstr\"{o}m black hole and the central singularity is replaced by a regular de Sitter core. We also show that if the previous EOS is replaced by a nonlocal one, it is possible to construct a self-gravitating fuzzy DM droplet but also in this case, the radial pressure is negative. Finally, we contemplate scenarios of different dark matter black holes with moderate mass values which could have formed in  galaxies. In particular, we probe the possibility whether such black holes could also be the central galactic objects.}}

\end{abstract}

\pacs{XXX}
\maketitle
\
\section{Introduction}
To examine the nature of Black Holes (BH) \cite{BH} more closely, it would be desirable to infer more about their interior structure, be it from the geometric point of view \cite{BHinterior1} or probing into the question what kind of matter has contributed to its formation \cite{BHinterior2}. In view of the fact that baryonic and leptonic matter  constitutes only four percent of the content of the universe and secondly recalling that the density profiles of the major component, Dark Matter (DM), grows as we approach the galactic center \cite{profiles}, the question whether the central galactic BH has a close connection to DM is a well-posed problem. Of course, the no-hair theorem \cite{nohair} prevents us from inferring observationally the inside properties of a BH, but modelling the interior structure of a BH (indeed, theoretical models of the interior BH are quite common in literature \cite{BHmodels2}) could reveal if a connection between DM and BH is feasible and at the same time be possibly a harbinger of new physics regarding both, the BH and the DM. An attempt in this direction has recently been proposed in \cite{Bosh} where the authors model the central galactic object assuming a DM profile fitted to the outer galactic region.  Other possible inter-connections between the two important components in the galactic bulge  have been examined in \cite{DMBH} where the authors consider the possibility of the growing of a BH by the DM absorption. This again suggests the BH as a seed for galactic structure. If so, it not unreasonable to think of a BH as made purely from DM. Motivated by this scenario we develop a new model of the galactic central object as a fuzzy BH (or droplet) in close analogy to BH/droplets inspired by non-commutative geometry \cite{BHnoncomm} where the Gaussian matter distribution and the de Sitter EOS play an important role. The above mentioned approach to BH physics can be generalized based on the fact that the Gaussian distribution is a special case of what is known as the Einasto profile of DM. This allows us to establish a possible connection between BH and DM, i.e., in constructing a fuzzy BH we follow the steps of a non-commutative BH with a new density profile, the Einasto parameterization. In this paper, we test the possibility of a  connection between the galactic BH and DM at the current stage of the Milky Way. However, in principle, we could also entertain the possibility of a smaller primordial fuzzy BH made out of DM and growing with time by absorbing matter and DM. We will leave such  a project to future undertakings.

Before the existence of a supermassive black hole at the centre of the Milky Way \cite{Ghez1,Ghez2,Nobel}, known as Sagittarius A$^*$, 
was widely accepted, there have been several attempts to construct theoretical models replacing the central black hole with other gravitational objects such as: gravastars \cite{MazMo,ChiRe}, bosonstars \cite{RuBo,ScMi}, naked singularities \cite{Joshi,BaMa,Chowd}, burning disks \cite{Kundt}, quantum cores (Ruffini-Arg\"{u}elles-Rueda model) \cite{Ru1,Ru2} and gravitationally bound clamps of dark matter relying on the exponential-sphere  density profile \cite{Sofue,Leu,Bosh}. It is worthwhile noticing that our information on the central galactic BH comes mainly from observing star orbits \cite{Ghez1,Ghez2}. In any model beyond the standard BH, it is obligatory to pay attention to the condition that the effective potential of the model, be it a droplet or a BH, resembles the standard BH effective potential, at least in the region of the bound orbits, i.e., around the local minimum of the potential (see, e.g,  \cite{Bosh}). In this respect, we will look for suitable parameters of the Einasto profile to model a fuzzy BH, a DM droplet or simply a BH made out of DM at the center of the galaxy in such a way that the observational data are confirmed by our model. Preferably, we will favour parameters already fitted to observational data, but we will also entertain the possibility that at the center of the galaxy the density profiles exhibits a different behaviour. To reach our goals, 
we will couple the aforementioned profile with an energy-momentum tensor for an anisotropic fluid and an equation of state of the form $p_r=-\rho$. Such an equation of state is quite common in the physics of BH \cite{BHmodels2}. In the present work, it leads to several regular BH models such that for each value of the parameters $\xi$ and $h$ entering in the Einasto profile, the mass parameter can be tuned so that a black hole will be present at the centre of a galaxy. This black hole has a horizon structure reminiscent of that we observe in the case of a Reissner-Nordstr\"{o}m geometry. Furthermore, the inner region of the black hole does not exhibit a curvature singularity at $r=0$ which is instead replaced by a regular de Sitter core. We also compute the Hawking temperature for the Einasto inspired black hole: we discover that the black hole increases its temperature, as the horizon radius shrinks, until the temperature reaches a maximum after which the temperature decreases sharply and vanishes exactly at the radius of the extremal black hole. But we will also demonstrate that a DM droplet would eventually result in an effective potential in accordance with observational data on the orbits. {\bf{In the second model, we consider a nonlocal equation of state for an anisotropic fluid. We observe that also in this case a negative pressure cannot be avoided. In particular, we construct a self-gravitating fuzzy DM droplet regular at the origin whose effective potential allows bound states for massive particle. Also in this case there is no central singularity at $r=0$.}}

The paper is organized as follows: In Section~\ref{Sec2}, we introduce the Einasto profile and some relevant formulae needed in the sections to follow. In Section~\ref{Sec4}, we investigate DM objects assuming a de Sitter-like EOS. {\bf{In section \ref{Sec5} we use a nonlocal equation of state to show that a fuzzy DM droplet made of an anisotropic fluid allows for stable orbits of massive particle. Finally, we present our conclusions in section~\ref{Sec6}.}}

\section{The Einasto density profile}\label{Sec2}
In his 1969 seminal paper \cite{Einasto}, Einasto showed that any realistic model aiming to give a faithful description of a galactic system should be characterized by certain descriptive functions such as the cumulative mass profile, the gravitational potential, the surface mass density etc., all satisfying a given set of constraints. Since such descriptive functions are integrals of the density profile $\rho=\rho(r)$ with $r$ a radial variable, it is natural to think that the most fundamental descriptive function of a galactic model is represented by the density profile itself which should exhibit the following properties
\begin{enumerate}
\item
$0<\rho(r)<\infty$ for all $r>0$;
\item
$\rho\in C^{\infty}(\mathbb{R}_+)$ with $\rho(r)\to 0$ as $r\to\infty$, i.e. it is a smooth and decreasing function that vanishes asymptotically at space-like infinity;
\item
certain moments associated to $\rho$ such as the central gravitational potential, the
total mass, and the effective radius of the system must be finite;
\item
the aforementioned descriptive functions must not exhibit jump discontinuities.
\end{enumerate}
Since then, the DM Einasto profile has been used not only to model several galaxies such as M31, M32, M87, Fornax and Sculptor dwarfs, and the Milky Way \cite{Einasto} but also to describe the density of dark matter haloes, see for instance \cite{Navarro,Springel,Mamon,Cardone2005,Mer,Ha,Gao,Dhar,Navarro2010,Chemin}. Regarding recent analytical studies of the Einasto model we refer to \cite{Cardone2005,Dhar,Retana}. 

We recall that the Einasto density profile \cite{Einasto}, which is generally adopted to describe cold DM halos \cite{Navarro,Mer,Gao,Ha,Potter,Navarro2010,deSalas} as well as the surface brightness of early-type galaxies and the bulges of spiral galaxies \cite{Davies,Caon,Don,Cell,Andre,Prug,Moll,Grah,Grah1,Gad}, is represented by the function
\begin{equation}\label{ein1}
\rho(r)=\rho_s\mbox{exp}\left(-d_\xi\left[\left(\frac{r}{r_{s}}\right)^{1/\xi}-1\right]\right),
\end{equation}
where $\xi$ is the Einasto index, $r_s$ the radius of the sphere enclosing half of the total mass, $\rho_s$ the mass density at $r=r_s$ and $d_\xi$ a numerical constant controlling that $r_s$ is indeed the half-mass radius. In the context of DM halos, the above density is also rewritten as \cite{deSalas}
\begin{equation}\label{ein2}
\rho(r)=\rho_{-2}\mbox{exp}\left(-2\xi\left[\left(\frac{r}{r_{-2}}\right)^{1/\xi}-1\right]\right),
\end{equation}
where $\rho_{-2}$ and $r_{-2}$ are the density and the radius at which the density profile behaves like $r^{-2}$, i.e. $d\ln{\rho}/d\ln{r}=-2$. If we introduce the central density
\begin{equation}
\rho_0=\rho_s e^{d_\xi}=\rho_{-2}e^{2\xi}
\end{equation}
and the scale length
\begin{equation}
h=\frac{r_s}{d_\xi^\xi}=\frac{r_{-2}}{(2\xi)^\xi}
\end{equation}
as in \cite{Retana}, it is straightforward to verify that the density profile becomes
\begin{equation}\label{ein3}
\rho(r)=\rho_0e^{-\left(\frac{r}{h}\right)^{1/\xi}}
\end{equation} 
and by adjusting the triple of parameters $\{\rho_0,h,\xi\}$, it is possible to model a variety of astrophysical objects. For instance, we have $4.54\lesssim \xi\lesssim 8.33$ for DM haloes with masses in the range of dwarfs to clusters \cite{Navarro}, $\xi\sim 5.88$ for galaxy-sized haloes \cite{Ha,Gao}, $\xi\sim 4.35$ for cluster-sized haloes in the Millenium Run \cite{Springel} and $\xi\sim 3.33$ for the most massive haloes for the Millenium Simulation \cite{Springel,Gao}. Since formula (\ref{ein3}) is equivalent to (\ref{ein1}) and (\ref{ein2}), there is no loss in generality if we work with the expression of the density profile given by (\ref{ein3}). Furthermore, the mass function $m$ and the gravitational potential $\Phi$ can be computed by solving the following ODEs obtained from the Newtonian equations of hydrostatic equilibrium, namely
\begin{equation}\label{HE}
\frac{dm}{dr}=4\pi r^2\rho(r),\quad
\frac{d\Phi}{dr}=\frac{G_N m(r)}{r^2},
\end{equation}
where $G_N$ denotes Newton's gravitational constant. As in \cite{Retana}, we immediately find that the total mass $M$ associated to the Einasto density profile is
\begin{equation}\label{massatotale}
M=4\pi\int_0^\infty x^2\rho(x)~dx=4\pi\rho_0 h^3 \xi\Gamma(3\xi),
\end{equation}
where $\Gamma$ denotes the Gamma function. The above relation allows to express the central density $\rho_0$ in terms of the total mass, and hence, we can rewrite (\ref{ein3}) as
\begin{equation}\label{dpfE}
\rho(r)=\frac{M}{4\pi h^3 \xi\Gamma(3\xi)}e^{-\left(\frac{r}{h}\right)^{1/\xi}}.
\end{equation}
Furthermore, a straightforward integration of the first equation in (\ref{HE}) leads to the following cumulative mass distribution
\begin{equation}\label{Mr}
m(r)=\frac{M}{\Gamma(3\xi)}\gamma\left(3\xi,\left(\frac{r}{h}\right)^{1/\xi}\right),
\end{equation}
where $\gamma$ denotes the lower incomplete Gamma function \cite{gamma}. {\bf{For a detailed analysis of the classical gravitational potential we refer to \cite{Retana}.}}

\section{Fuzzy black holes}\label{Sec4}
\begin{figure}[!ht]\label{hicsuntosemleones}
\includegraphics[scale=0.35]{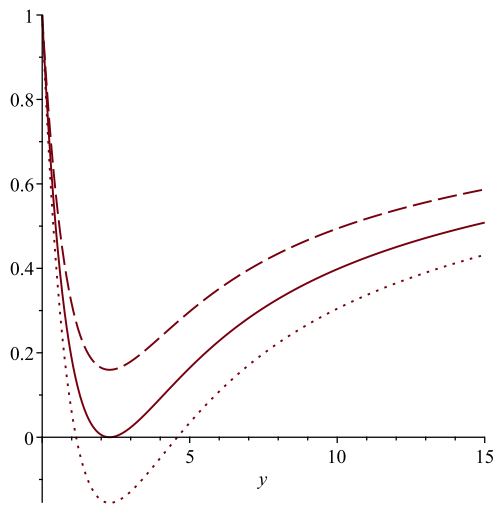}
\caption{\label{Bildchenstiri}
Plot of $g_{00}$ as a function of $y=(r/h)^{1/\xi}$ with $\xi=1/2$ for different values of $\mu$. The intersections on the horizontal axis represent the radii of the event horizons expressed in the variable $y$. If $\mu=\mu_0=0.95206$, there is one degenerate horizon at $y_0=2.28378$ (solid line). For $\mu=1.1>\mu_0$ (dot line), there are two horizons and in the case $\mu=0.8<\mu_0$ (dash line), there is no horizon. The latter case corresponds to a self-gravitating droplet consisting of an anisotropic fluid.}
\end{figure}

\begin{figure}[!ht]\label{hicsuntdevetleones}
\includegraphics[scale=0.35]{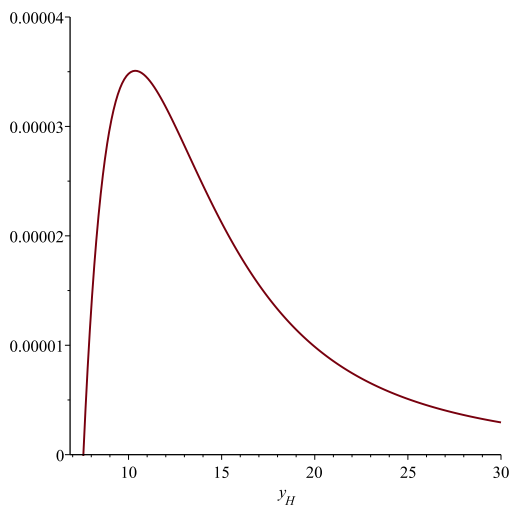}
\caption{\label{Bildchensest}
 Plot of $T_H$ versus $y_H=(r_H/h)^{1/\xi}$ in $h$ units for $\xi=3$. We have $T_H=0$ for $y_H=y_0=7.57496$, which coincides with the event horizon of the corresponding extremal black hole. The maximum temperature is $T_{H,max}= 3.5\cdot 10^{-5}$ and it corresponds to a mass parameter $\mu=787.66$.}
\end{figure}

In this section, we show that it is possible to construct black hole solutions from the Einasto density profile. Let us suppose that the mass density of a static, spherically symmetric, smeared gravitational source of total mass $M$ be modeled by the density profile (\ref{dpfE}) which contains as a special case ($\xi=1/2$ and $h=\sqrt{\theta}$) the Gaussian profile adopted by \cite{Piero} in the derivation of the noncommutative geometry inspired Schwarzschild black hole. Furthermore, we consider the following ansatz
\begin{equation}\label{metrik}
ds^2=g_{00}(r)dt^2-\frac{dr^2}{g_{00}(r)}-r^2\left(d\vartheta^2+\sin^2{\vartheta}d\varphi^2\right),\quad 0\leq\vartheta\leq\pi,\quad0\leq\varphi<2\pi
\end{equation}
representing a static, spherically symmetric manifold. We want to find the unknown function $g_{00}$ appearing in (\ref{metrik}) so that the above line element is a solution of the Einstein field equations coupled to the energy-momentum tensor of a static, anisotropic fluid and in the limit  
$r/h\to\infty$ such a solution goes over into the usual Schwarzschild metric. As in \cite{Piero} we introduce the energy-momentum tensor of a static, anistropic  fluid with density source (\ref{dpfE}) given by
\begin{equation}\label{enmomten}
T^\mu{}_\nu=\mbox{diag}(\rho,-p_r,-p_\bot,-p_\bot),\quad p_r\neq p_\bot,
\end{equation}
where $p_r$ and $p_\bot$ are the radial and tangential pressures, respectively, and we consider the Einstein field equations
\begin{equation}\label{EFGA}
R_{\mu\nu}=-8\pi\left(T_{\mu\nu}-\frac{T}{2}g_{\mu\nu}\right),\quad T=g^{\mu\nu}T_{\mu\nu}
\end{equation}
for the line element (\ref{metrik}). If we proceed as in \cite{Fliessbach}, we can use the conservation equation $T^{\mu\nu}{}_{;\nu}=0$ with $\mu=1$  in the $(\mu,\nu)=(2,2)$ equation coming  from (\ref{EFGA}) to obtain the Tolman-Oppenheimer-Volkoff equation, i.e. the general relativistic hydrostatic equilibrium equation given by
\begin{equation}\label{hydeqGl}
\frac{dp_r}{dr}+(\rho+p_r)\frac{m(r)+4\pi r^3 p_r}{r\left[r-2m(r)\right]}+\frac{2}{3}\left(p_r-p_\bot\right)=0,
\end{equation}
where the mass function is defined as
\begin{equation}\label{masse}
m(r)=4\pi\int_0^r u^2\rho(u)~du=\frac{M}{\Gamma(3\xi)}\gamma\left(3\xi,\left(\frac{r}{h}\right)^{1/\xi}\right).
\end{equation}
in the case of the Einasto profile. On the other hand, from the equation $T^{1\nu}{}_{;\nu}=0$  we get
\begin{equation}\label{druckGL}
-\frac{dp_r}{dr}=\frac{1}{2}g^{00}\frac{dg^{00}}{dr}(p_r+\rho)+\frac{2}{r}(p_r-p_\bot).
\end{equation}
and if we require that
\begin{equation}\label{pr}
p_r=-\rho=-\frac{M}{4\pi h^3 \xi\Gamma(3\xi)}e^{-\left(\frac{r}{h}\right)^{1/\xi}},
\end{equation}
then (\ref{druckGL}) can be solved for $p_\bot$ giving
\begin{equation}\label{druckT}
p_\bot=-\rho-\frac{r}{2}\frac{d\rho}{dr}=-\left[1-\frac{1}{2\xi}\left(\frac{r}{h}\right)^{1/\xi}\right]\rho.
\end{equation}
This procedure ensures that the conservation equation for the energy-momentum tensor is identically
satisfied. It is not difficult to verify that the tangential pressure vanishes at 
\begin{equation}\label{zeror}
r_0=(2\xi)^\xi h
\end{equation}
and takes its maximum value at
\begin{equation}\label{maxr}
r_m=\left(1+2\xi\right)^\xi h,\quad p_\bot(r_m)=\frac{M e^{-(2\xi+1)}}{8\pi h^3 \xi^2\Gamma(3\xi)}.
\end{equation}
Moreover, we have $p_r(0)=p_\bot(0)$.  Finally, we observe that equation (\ref{hydeqGl}) is trivially satisfied once $p_r$ and $p_\bot$ have been chosen as in (\ref{pr}) and (\ref{druckT}), respectively. This approach shows  that the Einasto matter distribution $\rho$ may describe a self-gravitating droplet of anistropic fluid.
\begin{figure}[!ht]\label{hicsuntsestleones}
\includegraphics[scale=0.35]{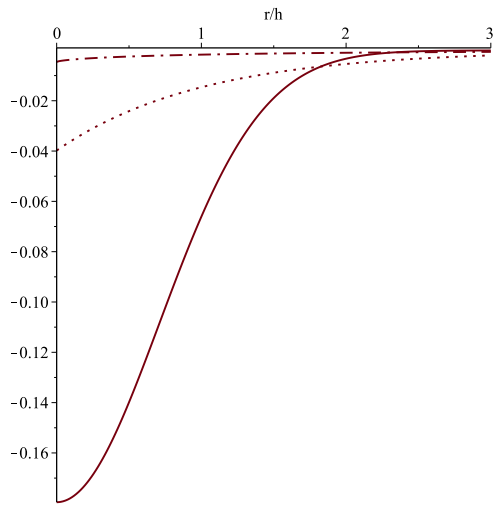}
\caption{\label{Bildchendva}
Plot of $h^3 p_r/M$ given by (\ref{pr}) for $\xi=1/2$ (solid line), $\xi=1$ (dotted line) and $\xi=1.5$ (dashdot line). In the DM case ( $\xi=7.072$, $h=2.121\cdot 10^{-9}$ Kpc and $M=4.57\cdot 10^{9} M_\odot$ \cite{Einasto}) we have $|p_r(0)|\approx 5.8\cdot 10^{-41}$ m$^{-2}$ (geometric units) or equivalently, $|p_r(0)|\approx 7\cdot 10^{3}$ $N/$m$^{2}$ (SI units).}
\end{figure}
\begin{figure}[!ht]\label{hicsuntsedamleones}
\includegraphics[scale=0.35]{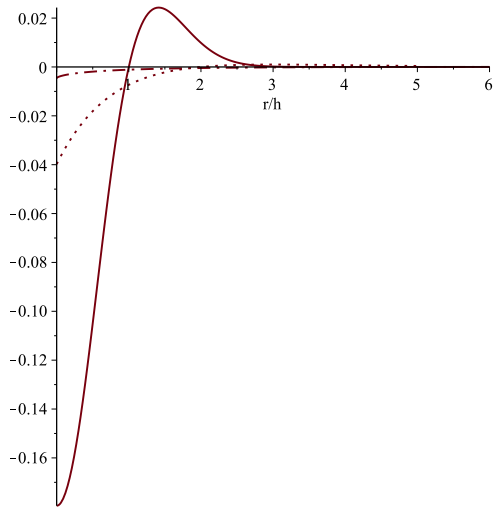}
\caption{\label{Bildchentri}
Plot of $h^3 p_\bot/M$  defined in (\ref{druckT}) for $\xi=1/2$ (solid line), $\xi=1$ (dotted line) and $\xi=1.5$ (dashdot line). In the DM ( $\xi=7.072$, $h=2.121\cdot 10^{-9}$ Kpc and $M=4.57\cdot 10^{9} M_\odot$ \cite{Einasto}) we have  $|p_\bot(0)|\approx 5.8\cdot 10^{-41}$ m$^{-2}$ (geometric units) or equivalently, $|p_\bot(0)|\approx 7\cdot 10^{3}$ $N/$m$^{2}$ (SI units). Note that $p_\bot$ changes sign at $r_0\approx 0.3$ Kpc which is situated well inside the Narrow Line Region of an Active Galactic Nuclei ($100$ pc$\div 4$ Kpc) \cite{Urry}..}
\end{figure}
If we consider the $(\mu,\nu)=(0,0)$ or equivalently, the $(\mu,\nu)=(1,1)$ equations in (\ref{EFGA}) together with (\ref{pr}) and the requirement that the metric goes over into the Minkowski metric asymptotically at infinity, we end up with the line
element
\begin{equation}\label{fmetric}
ds^2=\left(1-\frac{2m(r)}{r}\right)dt^2-\left(1-\frac{2m(r)}{r}\right)^{-1}dr^2-r^2\left(d\vartheta^2+\sin^2{\vartheta}d\varphi^2\right),
\end{equation}
where the mass function is given by (\ref{masse}). By means of $6.5.3$ in \cite{Abra} it is straightforward to verify that in the limit $r/h\to\infty$ the metric (\ref{fmetric}) reproduces the classic Schwarzschild metric. Furthermore, if we consider the $g_{00}$ component of the metric as a function of $r/h$ and we introduce the scaled mass $\mu=M/h$, it is possible to show that there exists a value of $\mu$, say $\mu_0$, such that $g_{00}$ has a double root at $x_0=r_0/h$. The numerical values of the extremal rescaled mass $\mu_0$ and the degenerate horizon $x_0$ for different values of  $\xi$ have been displayed in Table~\ref{tabler0}. They have been obtained by expanding the lower incomplete Gamma function in the expression for $g_{00}$ with the help of $6.5.29$ in \cite{Abra} where the first hundred terms in the expansion have been considered. Moreover, if  $\mu>\mu_0$ there exist two distinct horizons $r_1$ and $r_2$, and no horizon if $0<\mu<\mu_0$. Figure~\ref{Bildchenstiri} displays the plot of $g_{00}$ which exhibits the behaviours described above, i.e. $2$ horizons, $1$ horizon and no horizon.
\begin{table}[!ht]
\caption{Numerical values for the degenerate horizon $y_0=(r_0/h)^{1/\xi}$ and the corresponding extremal mass $\mu_0$ for different values of the Einasto parameter. The case $\xi=7.072$ corresponds to the DM case studied in \cite{Einasto}.}
\begin{center}
\begin{tabular}{ | l | l | l | l|l|}
\hline
$\xi$      &  $\mu_0$                 & $y_0$              \\ \hline
1/2      &  0.95206                 & 2.28378            \\ \hline
1        &  2.57470                 & 3.38364            \\ \hline
1.5      &  8.48079                 & 4.45141            \\ \hline
2        &  32.1069                 & 5.50210            \\ \hline
3        &  623.869                 & 7.57496            \\ \hline
4        &  16411.5                 & 9.62616            \\ \hline
5        &  5.43917$\cdot 10^{5}$   & 11.6647            \\ \hline
6        &  2.17332$\cdot 10^{7}$   & 13.6953            \\ \hline
7.072    &  1.34763$\cdot 10^{9}$   & 15.8657            \\ \hline
\end{tabular}
\label{tabler0}
\end{center}
\end{table}
Moreover, in the extreme and non extreme regimes, that is $\mu\geq\mu_0$, the behaviour of the metric coefficient $g_{00}$ for $r\to 0$ can be obtained by means of $6.5.29$ in \cite{Abra} and we find that 
\begin{equation}
g_{00}(r)\approx 1-\frac{2\mu}{\Gamma(3\xi+1)}\left(\frac{r}{h}\right)^2.
\end{equation}
The result above signalizes that differently as in the Schwarzschild black hole where there is a singularity at $r=0$, the central region is represented by a regular de Sitter core. Hence, the Einasto density profile coupled with an energy momentum tensor for an anisotropic fluid cures the problem of the central singularity provided that an EOS for the radial pressure is assumed to be of the form $p_r=-\rho$. Finally, in the case $\mu<\mu_0$ there is no event horizon {\bf{and also no naked singularity because the central region around the origin is described by a de Sitter geometry.}} We conclude this section by considering the Hawking temperature for this new class of black holes. The black hole temperature can be computed from the formula \cite{Piero}
\begin{equation}
T_H=\frac{1}{4\pi}\left.\frac{dg_{00}}{dr}\right|_{r=r_H}=\frac{1}{4\pi r_H}\left[
1-\frac{r^3_H e^{-(r/h)^{1/\xi}}}{\xi h^3\gamma(3\xi,(r/h)^{1/\xi})}\right],
\end{equation}
where $r_H$ represents the position of the event horizon and the total mass $M$ has been expressed in terms of $r_H$ by using the horizon equation $r_H=2m(r_H)$. Note that in the case $r_H/h\gg 1$ the expression above reproduces the usual result $T_H=(4\pi r_H)^{-1}$. The scenario emerging from Fig.~\ref{Bildchensest}, where we plotted the temperature $T_H$ as a function of $r_H$, is that an Einasto inspired black hole increases its temperature, as the horizon radius shrinks, until $T_H$ reaches a maximum after which $T_H$ decreases sharply and vanishes exactly at the radius of the extremal black hole, that is at $r_H=r_0$. Furthermore, in the case of an extreme black hole the Hawking temperature must be identically zero because the metric component $g_{00}$ has a double root at $r=r_0$. Hence, instead of observing a blow-up behaviour of the BH temperature, we find  that the evaporation process leads to a zero temperature extremal black hole whose final configuration is entirely controlled by the Einasto parameter $\xi$, the scale factor $h$ and the black hole mass. As already pointed out in \cite{DavidePiero}, a final configuration characterized by a finite temperature inhibits any relevant back reaction, i.e a self-interaction of the radiated energy with its source. This implies that our solution is stable versus back reaction and can describe the entire black hole life until the final configuration. The presence of an inner Cauchy horizon may be a source of concern, in the sense that the inner region of our  black holes is unstable, however one may proceed as in \cite{DavidePiero} to show the stability of the Einasto inspired black hole interior. At this point, a remark on nomenclature is in order. {\bf{If there is no horizon, we call the object a fuzzy droplet. If at least, one horizon develops, we name it a fuzzy BH.}}

{\bf We end this section with a remark on negative pressure, singularity theorems and the involved scales. For instance, it is tempting to attribute the negativity of pressure to some quantum effects which would limit the applicability of the scales one is using in a model.
First, we should notice that with this subject we touch the forefront of science and different interpretations exist in literature. For instance in \cite{Mazur} the possibility  of a phase transition to a negative pressure $p =-\rho$  EOS is considered as existing already in Einstein gravity. The inclusion of quantum theory would only would complete the picture, but is not a necessary ingredient. In such a picture the question of scales does not really arise. In \cite{Dvali} a quantum ``Macro-Quantumness'' is advocated
with the claim that the quantum effects for the macroscopic black holes are more important than suggested by means of a semi-classical
reasoning. It seems that one should treat the black hole as macroscopic quantum phenomenon and again no scales limitations arise. In short, the result is that the assumption that a black hole of macroscopic size can always be described classically leads to a contradiction and the classical description must break down on macroscopic scales. This was initially suggested in the seminal work of \cite{Dvali} by means of the so-called quantum $N$-portrait, which has been further developed in \cite{Dvali1,Dvali0,Dvali2,Dvali3,Dvali4}. Moreover, in \cite{Giddings} and \cite{Liberati} a long distance origin for Hawking radiation is considered and a ``quantum atmosphere'' assumed. This again would hint towards macroscopic quantum phenomena for any size black hole.

Inter alia, we would like to add one more possible interpretation.
Our precise EOS, $p_r=-\rho$ is coined according to the physics with a cosmological constant. In such a case, it is interesting to notice that the cosmological constant can violate the singularity theorems or the assumptions underlying them.  By this token, the regular black hole which we constructed in this section is not limited by scales due to the possible quantum origin of the EOS.  To substantiate our claim, we give another example which has to do with violation of singularity theorems in cosmology due to the cosmological constant $\Lambda$. We refer to \cite{Felten} where the authors show that for $\Lambda >\Lambda_{crit}$ there is no initial singularity. This violates the global cosmological singularity theorems. In particular, no quantum mechanics is involved. It is rather an effect of the cosmological constant.
We think a similar mechanism happens in the construction of regular black hole solutions under the de Sitter EOS coined after the physics with the cosmological constant. If so, the physics is again not really restricted by scales dictated by quantum mechanics. We can look at it from yet another point of view. In general, negative pressure is a concept taken seriously in physics
\cite{negativeP}.  As pointed out in \cite{negativeP} a negative pressure is not forbidden by the laws of thermodynamics. It is considered mostly in liquids and has an underlying mechanism, which is not necessarily attributed to quantum mechanics. Again we can argue that as
such it is not a local phenomenon limited to quantum mechanical scales. A quick comparison with regular black hole physics where a negative
pressure is used, tells us that we can start with a de Sitter EOS as done here and in \cite{BHmodels2} or it emerges naturally like in \cite{Mazur} where the effect is attributed to General Relativity. In the next section with will construct yet another model where a self-gravitating droplet emerges with a negative pressure.}

\subsection{The effective potential}

\begin{figure}[!ht]\label{grd}
\includegraphics[scale=0.35]{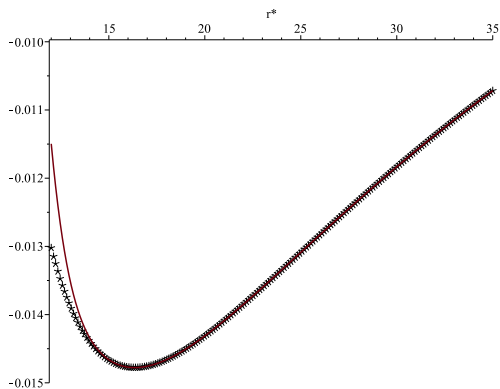}
\caption{\label{bildH10}
Plot of the effective potentials (\ref{VSmassive}) (asterisk symbol) (Schwarzschild case) {\bf{and (\ref{Vfull}) (solid line) (fuzzy droplet) in the massive case for $L=3$, $H=10$ and $\xi=0.2$. The free parameter $H$ is chosen so that it produces a scaling factor $h$ of the same order as the one predicted by \cite{Einasto}, i.e. $h_E=2.121\cdot 10^{-9}$ Kpc. The minimum of $V_{eff,S}$ is at $r^{*}_{min}=9+3\sqrt{6}\approx 16.35$ where $V_{eff,S}(r^{*}_{min})=-0.01477$ and $V_{eff}(r^{*}_{min})=-0.01477$. Both potentials share the same minimum and $V_{eff}$ is a good approximation of $V_{eff,S}$ in a neighbourhood of $r^{*}_{min}$ and asymptotically away.}}}
\end{figure}

\begin{figure}[!ht]\label{grd1}
\includegraphics[scale=0.35]{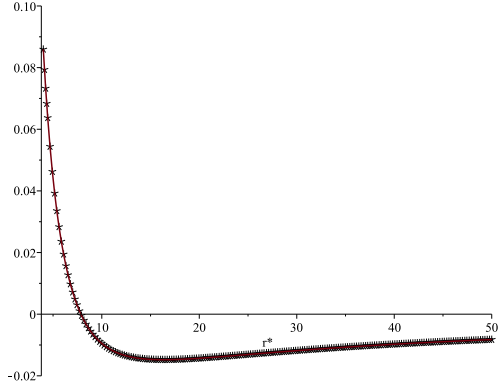}
\caption{\label{bildH01}
Plot of the effective potentials (\ref{VSmassive}) (asterisk symbol) (Schwarzschild case) and (\ref{Vfull}) (solid line) in the massive case when $L=3$, $H=0.1$ and $\xi=1.0$. This choice of $H$ gives rise to a scaling factor $h$ of the same order as $h_{B1}= 3.89\cdot 10^{-11}$ Kpc in \cite{deSalas}. The minimum is located at $r^{*}_{min}=9+3\sqrt{6}\approx 16.35$ where $V_{eff,S}(r^{*}_{min})=-0.0147$ and $V_{eff}(r^{*}_{min})=-0.0147$. Both potentials share the same minimum and $V_{eff,S}$ is a good approximation of $V_{eff}$ in a neighbourhood of the minimum and asymptotically away. }
\end{figure}

\begin{figure}[!ht]\label{gigi1}
\includegraphics[scale=0.35]{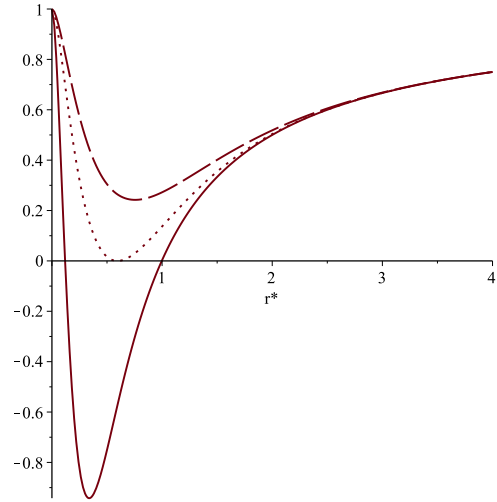}
\caption{\label{g00H01}
 Plot of the metric coefficient $g_{00}$ given by (\ref{g00c}) {\bf{in the case $H=0.1$ and $\mu=5$ for $\xi=1$ (solid line), $\xi=1.2865$ (dot line) and $\xi=1.4$ (longdashed line).}} Note that for $\xi=1$ the event horizon is at $r_*=1$ and coincides with that of a Schwarzschild black hole. The extreme black hole corresponds to $\xi=1.2865$ and its event horizon is located at $r^{*}_{e}=0.5947$ or equivalently at $r_{e}=2.33\cdot 10^{-7}$ pc.}
\end{figure}

If we insist that the black hole solution derived in Section~\ref{Sec4} sits at the centre of our galaxy, all observations should be the same. To this purpose, we study the problem whether our model of a diffuse dark matter black hole is able to fit the central galactic black hole in the Milky Way whose mass and Schwarzschild radius are  $M_{BH}=4.1\cdot 10^6~M_\odot$ and $R_{BH}=2G_N M_{BH}/c^2=17.4~R_\odot=3.92\cdot 10^{-7}$ pc, respectively \cite{Ghez1,Ghez2}. In order to do that, we need to find estimates for the Einasto parameter $\xi$ and the scaling factor $h$. This is done in two steps. First of all, we impose that the total mass $M$ entering in the line element (\ref{metrik}) through the metric coefficient $g_{00}$  coincides with $M_{BH}$. Secondly, we require that the mass function $m$ provides a good approximation for $M_{BH}$ when it is evaluated at the minimum $r_{min}$ of the Schwarzschild effective potential for a massive particle. More precisely, we exploit the freedom to force that
\begin{equation}\label{condizione}
1-\frac{m(r_{min})}{M_{BH}}\leq 10^{-2}.
\end{equation}
In the analysis to follow, it is convenient to rewrite the above condition in the equivalent form
\begin{equation}\label{conditio1}
\Delta\gamma:=\frac{1}{\Gamma(3\xi)}\gamma\left(3\xi,\left(\frac{r_{min}}{h}\right)^{1/\xi}\right)-0.99\geq 0,
\end{equation}
where we made use of (\ref{masse}). As we will soon realize, it will turn out that the above condition not only ensures that the Schwarzschild effective potential and the effective potential of our diffused gravitational object share the same minimum but they both also agree in a large neighbourhood of it and asymptotically away (see for instance Fig.~\ref{bildH10}). 
The effective potential $V_{eff}$ for the problem at hand is obtained  from the geodesic equation. More precisely, following the same procedure as in \cite{Fliessbach}, we can bring the radial equation into a form of an energy conservation equation, namely
\begin{equation}
\frac{\dot{r}^2}{2}+V_{eff}(r)=const,
\end{equation}
where the dot means differentiation with respect to the proper time or an affine parameter, depending whether we consider the case of a massive or a  massless particle. By means of equation (25.27) in \cite{Fliessbach} we immediately find that the effective potential associated to the geometry described by the line element (\ref{metrik}) is
\begin{equation}\label{realmassive}
V_{eff}(r)=\frac{\ell^2}{2r^2}-\frac{M_{BH}}{\Gamma(3\xi)}\left(\frac{\epsilon}{r}+\frac{\ell^2}{r^3}\right)\gamma\left(3\xi,\left(\frac{r}{h}\right)^{1/\xi}\right),\quad\epsilon=\left\{
\begin{array}{cc}
1 &\mbox{if}~m_p\neq 0,\\
0 &\mbox{if}~m_p=0,
\end{array}
\right.
\end{equation}
where $m_p$ denotes the mass of a test particle and $\ell$ is its total angular momentum per unit mass. At this point, it is also useful to recall that the effective potential in the case of the Schwarzschild metric is
\begin{equation}\label{massive}
V_{eff,S}(r)=\frac{\ell^2}{2r^2}-M_{BH}\left(\frac{\epsilon}{r}+\frac{\ell^2}{r^3}\right).
\end{equation}
Let $r_s=2M_{BH}$. If we rescale the radial variable and the angular momentum per unit mass as $r^*=r/r_s$ and $L=\ell/r_s$, the Schwarzschild effective potential in the massive case becomes
\begin{equation}\label{VSmassive}
V_{eff,S}(r^*)=-\frac{1}{2r^*}+\frac{L^2}{2{r^{*}}^2}-\frac{L^2}{2{r^*}^3}
\end{equation}
and the event horizon is now located at $r_*=1$. Furthermore, it exhibits a minimum and a maximum at 
\begin{equation}
r^{*}_{min}=\frac{L^2}{2}\left(1+\sqrt{1-\frac{3}{L^2}}\right),\quad
r^{*}_{max}=\frac{L^2}{2}\left(1-\sqrt{1-\frac{3}{L^2}}\right)
\end{equation}
provided that $L>\sqrt{3}$. Introducing the same rescaling for (\ref{conditio1}) and for the effective potential (\ref{realmassive}) in the massive case yields
\begin{equation}\label{Vfull}
V_{eff}(r^*)=\frac{L^2}{2{r^*}^2}-\frac{1}{\Gamma(3\xi)}\left(\frac{1}{2r^*}+\frac{L^2}{2{r^*}^3}\right)\gamma\left(3\xi,\left(\frac{r^*}{H}\right)^{1/\xi}\right),\quad H=\frac{h}{r_s}
\end{equation}
and
\begin{equation}\label{conditio2}
\Delta\gamma:=\frac{1}{\Gamma(3\xi)}\gamma\left(3\xi,\left(\frac{r^{*}_{min}}{H}\right)^{1/\xi}\right)-0.99\geq 0.
\end{equation}
The above condition is an inequality in the free parameters $H$ and $\xi$. To show that its solution set is non empty, we will first consider different choices of $H$ so that the corresponding scale factors $h=r_s H$ have the same orders of magnitude of the scaling factors appearing in \cite{Einasto} and \cite{deSalas}. For each choice of $H$ we solve the inequality (\ref{conditio2}) with respect to the parameter $\xi$. Since the particular value of $r^{*}_{min}$ depends on the rescaled total angular momentum $L$, the procedure outlined above requires that we also fix $L$. For instance, in \cite{Einasto} the scaling factor for a DM halo is $h_E=2.121\cdot 10^{-9}$ Kpc. Hence, if we choose $H=10$ the corresponding scaling factor in our model is $h=3.92\cdot 10^{-9}$ Kpc. To find out which values of $\xi$ will satisfy (\ref{conditio2}), we consider different values of $L$ and $r^{*}_{min}$. In the case $L=2$ and $r^{*}_{min}=6$, we find numerically that $\Delta\gamma<0$ in the range $10^{-6}\leq \xi\leq 13$ signalizing that the inequality (\ref{conditio2}) cannot be satisfied. The situation changes if we increase the value of $L$. If $L=3$ with $r^{*}_{min}=9+3\sqrt{6}$, it turns out that $\Delta\gamma<0$ for $\xi<0.32$. If $L=5$ with $r^{*}_{min}=25+5\sqrt{22}$, any $\xi<0.80$ will do the job while for $L=100$ with $r^{*}_{min}=10^4+10^2\sqrt{9997}$ it is necessary that $\xi<2.73$. Does our model predict a fuzzy BH or a fuzzy droplet  when $H=10$ and $\xi$ is chosen so that (\ref{conditio2}) is satisfied? To answer this question, we observe that in geometric units $r_s=2M_{BH}$ so that $M_{BH}=r_s/2$ and the rescaled mass parameter $\mu$ entering in our model will be fixed according to
\begin{equation}
\mu=\frac{M_{BH}}{h}=\frac{r_s}{2h}=\frac{1}{2H},
\end{equation}
where in the last step we used the relation $h=r_s H$. Moreover, by means of the rescaling $r^*=r/r_s$ together with the expansion $6.5.29$ in \cite{Abra} we can rewrite the metric coefficient $g_{00}$ according to
\begin{equation}\label{g00c}
g_{00}(r^*)=1-2\mu\left(\frac{r^*}{H}\right)^2e^{-(r^{*}/H)^{1/\xi}}\sum_{k=0}^\infty\frac{(r^{*}/H)^{k/\xi}}{\Gamma(3\xi+k+1)}.
\end{equation}
\begin{figure}[!ht]\label{g}
\includegraphics[scale=0.35]{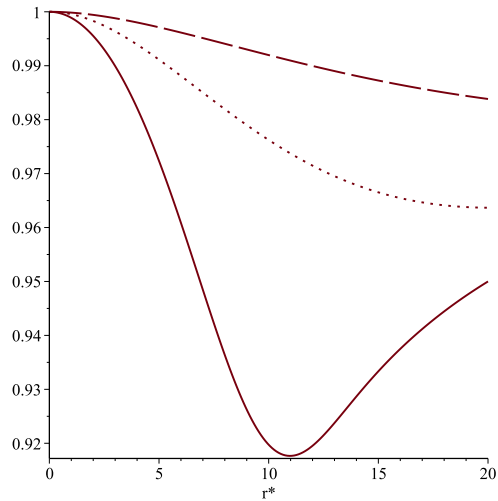}
\caption{\label{figg}
 Plot of the metric coefficient $g_{00}$ given by (\ref{g00c}) in the case $H=10$ and $\mu=0.05$ for $\xi=0.2$ (solid line), $\xi=0.7$ (dotted line) and $\xi=1.0$ (longdashed line).}
\end{figure}
As we can see from Fig.~\ref{figg}, the equation $g_{00}(r^*)=0$ does not admit any real root and therefore, this model predicts a fuzzy droplet.  More recent estimates of the DM density using the rotation curve of the Milky Way using different Galactic mass models together with certain DM and baryonic density distributions has been obtained by \cite{deSalas}.  In particular, \cite{deSalas} studied two baryonic models: the model B1 where a combination of Plummer's and Miyamoto-Nagai's potentials has been assumed and the model B2 which relies on the assumption of two different double exponential profiles and the Hernquist profile. Three different DM spherical halos were tested where one is described by  the  Einasto  profile. In  the  B1  model coupled to the  Einasto  profile, the scaling factor is $h_{B1}= 3.89\cdot 10^{-11}$ Kpc while the B2 model  predicts $h_{B2}=1.426\cdot 10^{-5}$ Kpc. Let us first consider the B1 model. In this case, we need to fix $H=0.1$ so that the scaling factor predicted by our model is of the same order as $h_{B1}$. More precisely, we have $h=3.92\cdot 10^{-11}$ Kpc. Proceeding as before, we find that, in order for (\ref{conditio2}) to be satisfied, $\xi<1.67$ for $L=2$, $\xi<1.98$ for $L=3$, $\xi<2.31$ for $L=5$ and $\xi<3.98$ for $L=100$. As it can be seen in Fig.~\ref{bildH01}, we observe that (\ref{Vfull}) is well approximated by the Schwarzschild effective potential in a large neighbourhood of the minimum also in the case $H=0.1$. To the value $H=0.01$ there corresponds a rescaled mass parameter $\mu=5$. A close inspection of Fig.~\ref{g00H01} shows that differently as in the case $H=10$ we have a more complex scenario. If $\xi<1.2865$, we have a dark matter black hole with two distinct horizons while for $\xi=1.2865$ an extreme black hole with radius $r_{e}=2.33\cdot 10^{-7}~\mbox{pc}< R_{BH}=3.92\cdot 10^{-7}~\mbox{pc}$. Finally, if $\xi>1.2865$, there is a diffused dark matter droplet without horizon. The same scenario occurs if we further reduce the value of $H$. For instance, if $H=0.02$, the extreme value of the Einasto parameter discriminating between a fuzzy black hole and a fuzzy droplet is found to be $\xi=1.9093$. Regarding the B2 model in \cite{deSalas}, it is necessary to fix $H=10^5$. It turns out that it is not possible to find any value of the parameter $\xi$ such that the inequality (\ref{conditio2}) is satisfied. This means that the matching procedure at the minimum of the effective potential which ensures at $r=r_{min}$  that the mass function approximates $M_{BH}$ according to the condition (\ref{condizione}) cannot be applied.    
\begin{figure}[!ht]\label{grdiz}
\includegraphics[scale=0.35]{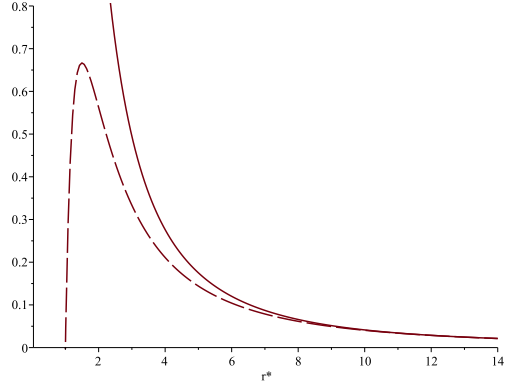}
\caption{\label{bildZ}
Plot of the effective potentials (\ref{VSmassive}) (long-dash line) (Schwarzschild case) and (\ref{Vfull}) (solid line) in the massless case when $L=3$, $H=10$ and $\xi=0.2$. This choice of $H$ gives rise to a scaling factor $h$ of the same order as $h_E=2.121\cdot 10^{-9}$ Kpc in \cite{Einasto} and central object is modelled in terms of a droplet (see also Fig.~\ref{figg}). The maximum of $V_{eff,S}$ is located at the radius of the photon sphere $r^{*}_{\gamma}=3/2$ while the event horizon of the Schwarzschild black hole is $r^{*}_{s}=1$. Note that the effective potential given by (\ref{Vfull}) does not exhibit a maximum and therefore, the droplet does not possess a photon sphere.}
\end{figure}
\begin{figure}[!ht]\label{grd10}
\includegraphics[scale=0.35]{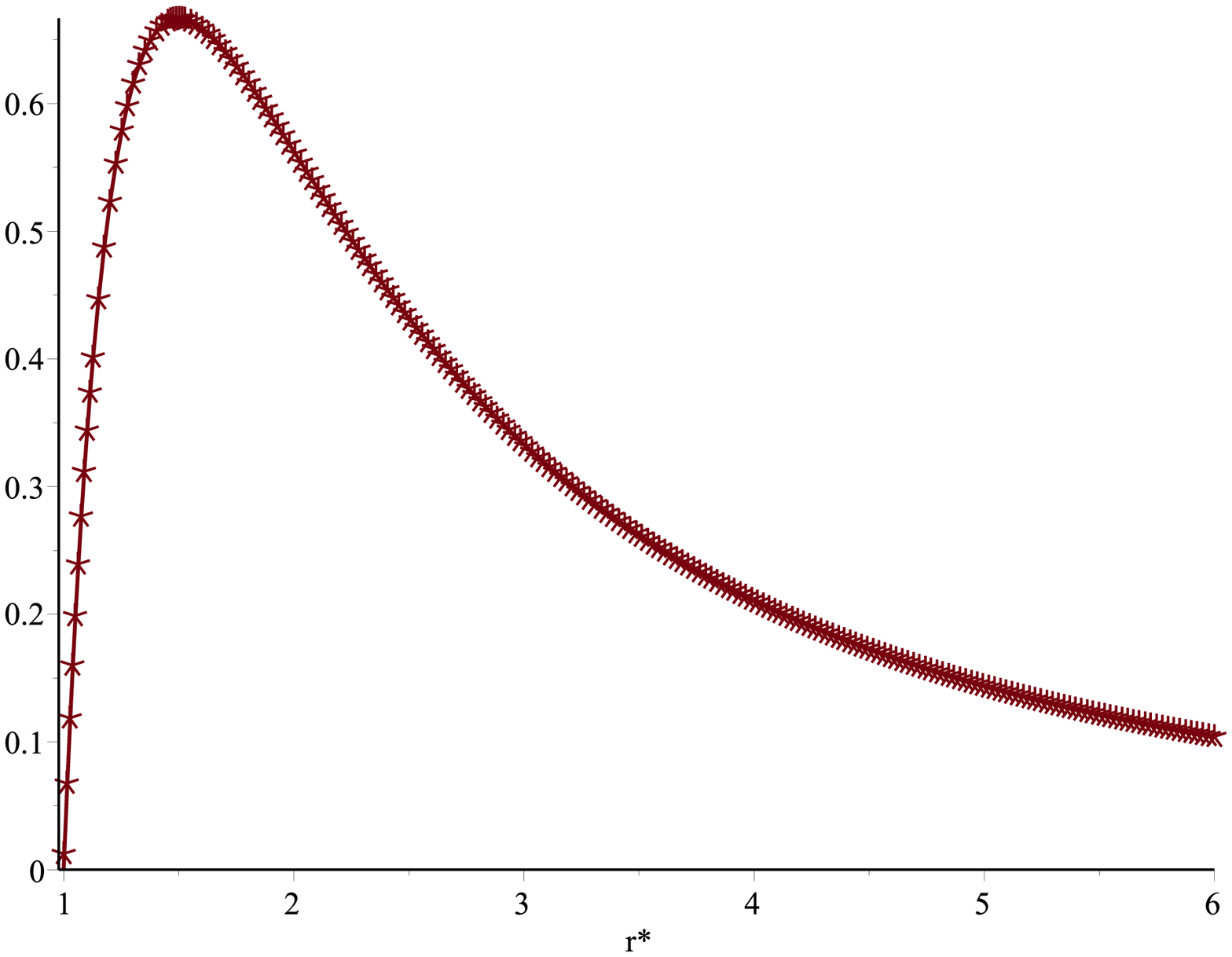}
\caption{\label{bildH010}
Plot of the effective potentials (\ref{VSmassive}) (asterisk symbol) (Schwarzschild case) and (\ref{Vfull}) (solid line) in the massless case when $L=3$, $H=0.1$ and $\xi=1.0$. Both potentials share the same photon sphere at $r^{*}_{\gamma}=3/2$. Both black hole models have the same event horizon at $r^{*}_{s}=1$.}
\end{figure}

\section{Diffused self-gravitating dark matter droplets from a nonlocal equation of state}\label{Sec5}
\begin{figure}[!ht]\label{hicsuntego}
\includegraphics[scale=0.35]{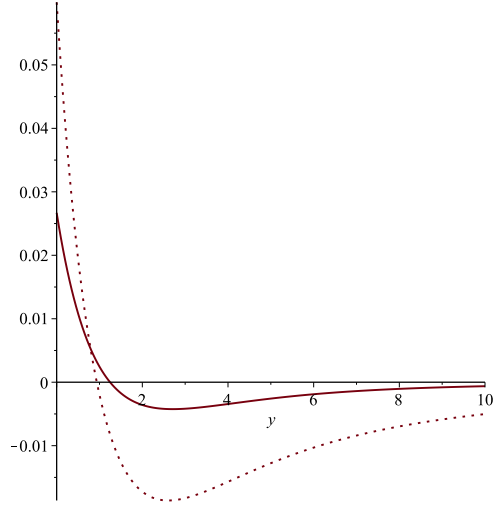}
\caption{\label{Bildchendevet}
 Plot of the rescaled radial pressure $h^2 p_r/\mu$ given by (\ref{KKL}) with $k=100$ versus $y=(r/h)^{1/\xi}$ in the case $\xi=1/2$ (dotted line) and $\xi=0.8$ (solid line). The radial pressure is positive in the inner region and it vanishes at some typical value of $y$ which depends on the particular choice of the Einasto parameter $\xi$. Outside such a value of $y$, the pressure becomes negative and it exhibits a minimum. The same behaviour can be observed for all other values of $\xi$ considered in Table~\ref{tablerx}.}
\end{figure}

\begin{figure}[!ht]\label{hicsuntte}
\includegraphics[scale=0.35]{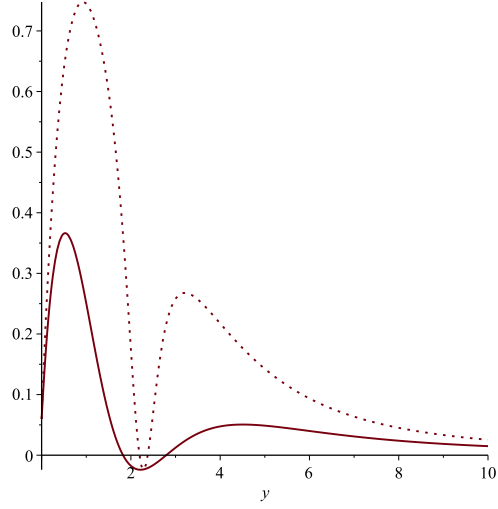}
\caption{\label{figurina}
 Plot of the tangential pressure $h^2p_\bot/\mu$ given in (\ref{ttp1}) versus $y=(r/h)^{1/\xi}$ in the case $\xi=1/2$ and for different values of the rescaled mass $\mu<\mu_0=0.95206$. The solid and dotted lines correspond to the cases $\mu=0.7$ and $\mu=0.94$, respectively. The plot has been obtained by applying $6.5.29$ in \cite{Abra} to expand the lower incomplete Gamma function in (\ref{ttp1}) where the first hundred terms in the expansion have been considered. The same behaviour can be observed for all other values of $\xi$ considered in Table~\ref{tablerx}.}
\end{figure}

\begin{figure}[!ht]\label{kontovel}
\includegraphics[scale=0.35]{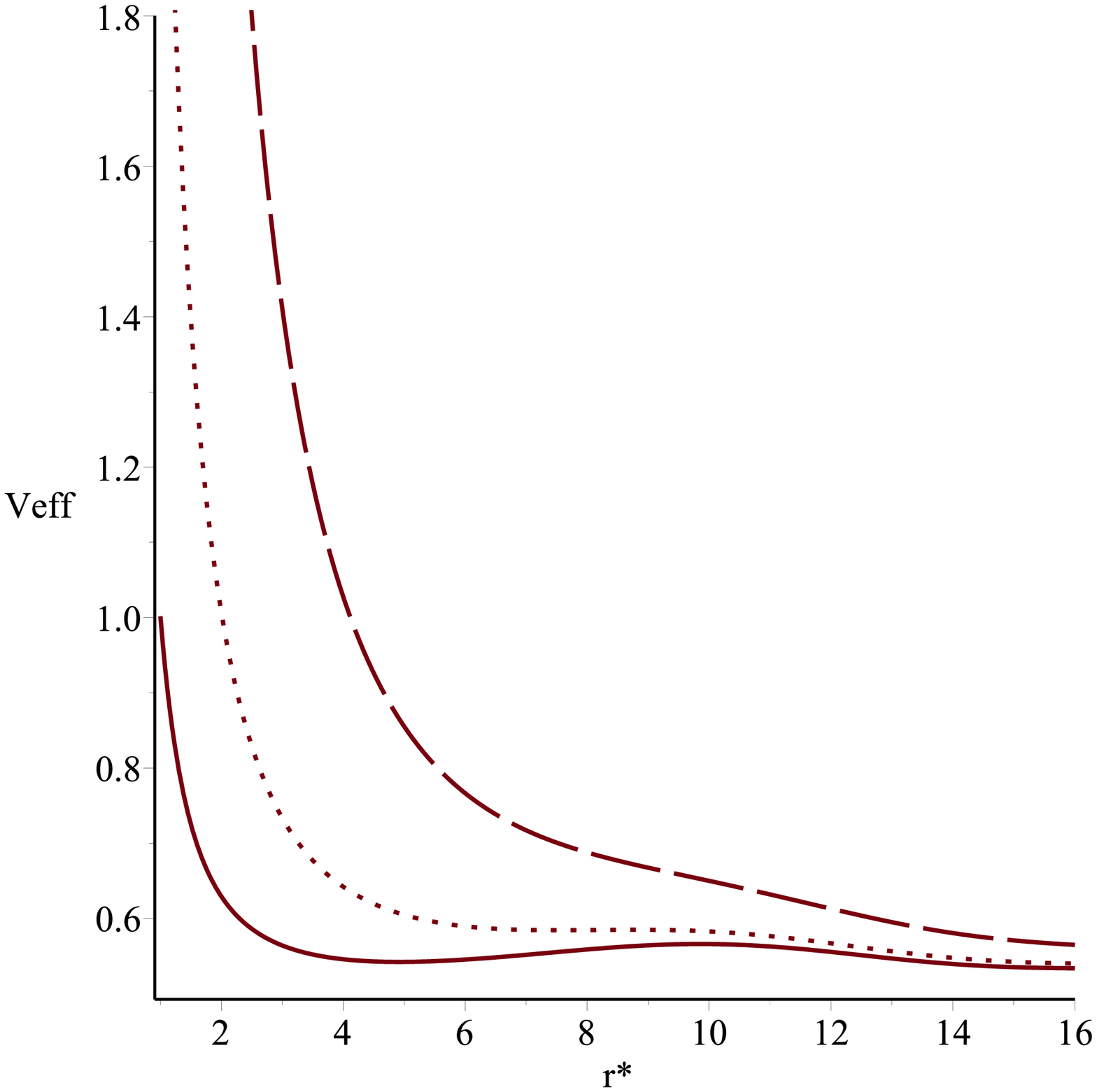}
\caption{\label{mlaka}
 {\bf{Plot of the effective potential (\ref{effp}) for $\epsilon=1$, $\xi=0.2$, $H=10$ with $L=1$ (solid line), $L=2$ (dotted line) and $L=5$ (long dashed line).}}}
\end{figure}

In the previous section, we assumed an equation of state for the radial pressure of the form, $p_r = -\rho$, and an anisotropic fluid with an additional tangential pressure because, if we would have insisted on a hydrostatical equilibrium, expressed through the Tolman-Oppenheimer-Volkov (TOV) equation, and an energy-momentum tensor of a perfect fluid, this approach would have led to an over-determined system of equations as the density $\rho$ is already assigned a priori and the pressure is fixed by the EOS. This allowed to show that starting with the Einasto density profile, it is possible to construct a fuzzy black hole or a diffused self gravitating droplet depending on the particular value of the rescaled mass parameter $\mu$. {\bf{In the present section, we offer a further example through a different EOS signalizing that the regularity of a fuzzy black hole or a fuzzy  self-gravitating droplet seems to require as a main feature that the radial pressure has to be negative at least on a subset of the positive real line. To this purpose, we need to fix a certain EOS and an energy-momentum tensor. Since the Einasto energy density has a diffused profile, we would expect that any change in the radial pressure should take into account the effects of the variations of the energy density within the entire volume. Hence, it seems reasonable to think that nonlocal effects may play a certain role when we work with such a distribution. For instance, in order to take into account nonlocality, we could imagine as in \cite{Hern1,Hern2,Ab1} that the components of the energy-momentum tensor besides displaying a dependence on the spacetime event it also exhibits a functional dependence by averaging the energy density over the enclosed configuration . Without further ado, let us}} derive the complete solution of the gravitational field equations for an Einasto  inspired anisotropic fluid described by a nonlocal equation of state of the form \cite{Hern1,Hern2,Ab1}
\begin{equation}\label{prrr}
p_r(r)=\rho(r)-\frac{2}{r^3}\int_0^r u^2\rho(u)~du=\frac{M}{4\pi\Gamma(3\xi)}\left[\frac{e^{-\left(\frac{r}{h}\right)^{1/\xi}}}{\xi h^3}-\frac{2}{r^3}\gamma\left(3\xi,\left(\frac{r}{h}\right)^{1/\xi}\right)\right].
\end{equation}
Since we are interested in matter configurations at hydrostatic equilibrium, {\bf{we can introduce an effective size}} $R$ of the object by the condition $p_r(R)=0$. The numerical value of $R$ can be found by plotting $h^2 p_r$ versus $y=(r/h)^{1/\xi}$. To this purpose, it is convenient to introduce the mass parameter $\mu=M/h$ so that (\ref{prrr}) becomes
\begin{equation}\label{KKL}
\frac{h^2 p_r}{\mu}=\frac{1}{4\pi\Gamma(3\xi)}\left[\frac{e^{-y}}{\xi}-\frac{2}{y^{3\xi}}\gamma(3\xi,y)\right]
=\frac{e^{-y}}{4\pi}\left[\frac{1}{\xi\Gamma(3\xi)}-2\sum_{k=0}^\infty\frac{y^k}{\Gamma(k+3\xi+1)}\right],
\end{equation}
{\bf{where in the last step we made use of $6.5.29$ in \cite{Abra}. For a list of numerical values of $R^{*}=R/h$ we refer to Table~\ref{tablerx}. Moreover, in Figure~\ref{Bildchendevet} we plot the radial pressure to explicitly demonstrate that it is indeed positive in a region of finite size $R$ but negative outside. This however does not mean that the gravitational object will have a finite radius because the energy density does not vanish in the region $r>R$.}}
\begin{table}[!ht]
\caption{{\bf{For different choices of the parameter $\xi$ we present some typical values of the radial distance $R^{*}=R/h$ at which the radial pressure vanishes.  The case $\xi=7.072$ corresponds to the Dark Matter case studied in \cite{Einasto}. The numerical values has been obtained by setting equal to zero the expression in the brackets appearing in (\ref{KKL}) and choosing $k=100$. Note that $R^{*}$ grows linearly in the parameter $\xi$.}}}
\begin{center}
\begin{tabular}{ | l | l | l | l|l|}
\hline
$\xi$      &  $R^{*}=R/h$                           \\ \hline
1/2      &  0.93675                             \\ \hline
1        &  1.45123                             \\ \hline
1.5      &  1.95996                             \\ \hline
3        &  3.47355                             \\ \hline
4        &  4.47832                             \\ \hline
5        &  5.48161                             \\ \hline
6        &  6.48401                             \\ \hline
7.072    &  7.55797                             \\ \hline
\end{tabular}
\label{tablerx}
\end{center}
\end{table}
{\bf{In order to proceed further}}, we consider also in this case a spherically symmetric static matter distribution represented by the Einasto density profile $\rho$ but differently as in the previous section we now assume the following ansatz for the line element
\begin{equation}\label{guappo}
ds^2=A^2(r)dt^2-\frac{dr^2}{B(r)}-r^2\left(d\vartheta^2+\sin^2{\vartheta}d\varphi^2\right).
\end{equation}
If we suppose that the energy-momentum tensor in the region occupied by the matter distribution is again given by (\ref{enmomten}), the Einstein field equations $G_{\mu\nu}=-8\pi T_{\mu\nu}$ together with the conservation equation $T^{\mu\nu}{}_{;\nu}=0$ with $\mu=r$ give rise to the following coupled system of ODEs
\begin{eqnarray}
\frac{1}{r}\frac{dB}{dr}-\frac{1-B(r)}{r^2}&=&-8\pi\rho(r),\label{28}\\
\frac{2B(r)}{rA(r)}\frac{dA}{dr}-\frac{1-B(r)}{r^2}&=&8\pi p_r(r),\label{29}\\
\frac{1}{2rA(r)}\left[2B(r)\frac{dA}{dr}+2rB(r)\frac{d^2 A}{dr^2}+A(r)\frac{dB}{dr}+r\frac{dA}{dr}\frac{dB}{dr}\right]&=&8\pi p_\bot(r),\label{30}\\
\frac{dp_r}{dr}+\frac{p_r(r)+\rho(r)}{A(r)}\frac{dA}{dr}&=&\frac{2}{r}\left[p_\bot(r)-p_r(r)\right].\label{31}
\end{eqnarray}
With the help of equations (\ref{28}) and (\ref{29}) it is straightforward to check that equations (\ref{30}) and(\ref{31}) represent the same equation. Hence, we will restrict our attention to the differential system given by (\ref{28}), (\ref{29}) and (\ref{31}) and in order to avoid to work with an under-determined system (there are three equations for the four unknown functions $A$, $B$, $p_r$ and $p_\bot$), we also assume an equation of state for matter represented by (\ref{prrr}). Integrating equations  (\ref{28}) (\ref{29}) yields
\begin{eqnarray}
B(r)&=&1-\frac{2m(r)}{r},\label{Bfun}\\
A^2(r)&=&e^{\phi(r)},\quad\phi(r)=\int \psi(r)~dr,\quad\psi(r)=\frac{1}{B(r)}\left[8\pi rp_r(r)+\frac{2m(r)}{r^2}\right].\label{Afun}
\end{eqnarray}
where the mass function $m$ is given by (\ref{masse}), {\bf{while the tangential pressure $p_\bot$ is obtained directly from (\ref{31}) together with (\ref{Afun}), that is}}
\begin{equation}\label{ttp}
p_\bot(r)=p_r(r)+\frac{r}{2}\left[\frac{dp_r}{dr}+\frac{p_r(r)+\rho(r)}{B(r)}\left(4\pi rp_r(r)+\frac{m(r)}{r^2}\right)\right].
\end{equation}
{\bf{At this point a remark is in order. First of all, we observe that the metric coefficient $B$ is the same as the metric coefficient $g_{rr}$ appearing in the line element (\ref{fmetric}). This implies that the same analysis of the zeroes of $g_{rr}$ performed in Section~\ref{Sec4} applies to the present case as well. Moreover, $B$ appears in (\ref{ttp}) in the denominator and this will cause the tangential pressure to become singular at the zeroes of $B$. On the other hand, the line element (\ref{guappo}) can be cast into the form}}
\begin{equation}\label{guappofinal}
ds^2=e^{\phi(r)}dt^2-\left(1-\frac{2m(r)}{r}\right)^{-1}dr^2-r^2\left(d\vartheta^2+\sin^2{\vartheta}d\varphi^2\right),
\end{equation}
{\bf{which is reminiscent of a dirty black hole metric provided it satisfies the conditions formulated in \cite{PIEROBOSS}. Under the assumption of an anisotropic energy-momentum tensor, one of them is that $p_\bot$ remains finite. The other one requires $B$ to have two zeros. We will leave the examination if such a dirty black hole is viable at all to future projects and circumvent the latter condition by requiring  $\mu<\mu_0$ . Note that such a condition also guarantees that the function $\phi$ is everywhere regular because it prevents the function $B$ entering in (\ref{Afun}) from having real roots. Hence, we conclude that the line element (\ref{guappofinal}) describes a fuzzy self-gravitating dark matter droplet. In Figure~\ref{figurina} we plot the tangential pressure to show that it is indeed well-behaved for any value of $r$ provided that $\mu<\mu_0$. To accomplish that, it is convenient to introduce the variable $y=(r/h)^{1/\xi}$ which allows to rewrite (\ref{ttp}) as}}
\begin{equation}\label{ttp1}
\frac{h^2 p_\bot}{\mu}=\frac{1}{4\pi\Gamma(3\xi)}\left[\frac{\gamma(3\xi,y)}{y^{3\xi}}-\frac{ye^{-y}}{2\xi^2}\right]+\frac{\mu y^{2\xi}}{\Gamma^2(3\xi)\left[1-\frac{2\mu}{y^\xi\Gamma(3\xi)}\gamma(3\xi,y)\right]}\left[\frac{e^{-y}}{\xi}-\frac{\gamma(3\xi,y)}{y^{3\xi}}\right]^2.
\end{equation}  
\begin{figure}[!ht]\label{crinale}
\includegraphics[scale=0.35]{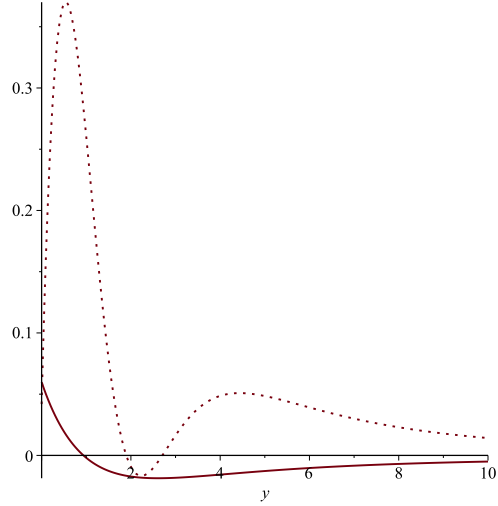}
\caption{\label{mocco}
 Plot of the radial and tangential pressures $h^2 p_r$ (solid line) and  $h^2p_\bot$ (dotted line) versus $y=(r/h)^{1/\xi}$ in the case $\xi=1/2$ and  $\mu=0.7$. $6.5.29$ in \cite{Abra} has been used to expand the lower incomplete Gamma function in (\ref{ttp1}) where the first hundred terms in the expansion have been considered. The minima of the radial and tangential pressures denoted by $y_r$ and $y_\bot$ do not coincide and they are located at $y_r=2.6038$ and $y_\bot=2.2335$, respectively.}
\end{figure}
{\bf{In Figure~\ref{mocco} we plot the radial and tangential pressures for the same choice of the Einasto parameter and the rescaled mass.}} Also in the case of a nonlocal equation of state, it turns out that the self-gravitating droplet does not exhibit any singularity at $r=0$. To verify that, we first observe that the Kretschmann scalar associated to the line element (\ref{guappofinal}) is given by
\begin{equation}
R^{\alpha\beta\gamma\delta}R_{\alpha\beta\gamma\delta}=\frac{2}{r^2}\left[\left(\frac{dB}{dr}\right)^2+B^2(r)\psi^2(r)\right]+\frac{1}{4}\left[B(r)\psi^2(r)+2B(r)\frac{d\psi}{dr}+\psi(r)\frac{dB}{dr}\right]^2.
\end{equation}
{\bf{Due to the presence of the term $1/r^2$ in the above expression, it is not clear if the Kretschmann scalar is singularity free. However, if we expand the lower incomplete Gamma function entering in the metric coefficients by means of  $6.5.29$ in \cite{Abra} and we let $r\to 0$, we find that}}
\begin{equation}
\lim_{r\to 0}R^{\alpha\beta\gamma\delta}R_{\alpha\beta\gamma\delta}=\frac{32M^2}{9h^6 \xi^2\Gamma^2(3\xi)}+\frac{12 M^2[(1+2\xi)\Gamma^2(3\xi)+\xi^2]}{9h^6 \xi^4\Gamma^4(3\xi)}
\end{equation}
{\bf{showing that there is no curvature singularity at $r=0$. Regarding the asymptotic behaviour of the line element (\ref{guappofinal}) we clearly have $B\to 1$ at space-like infinity while by means of $6.5.3$ and $6.5.32$ in \cite{Abra} it is possible to show that the function $\phi$ exhibits the asymptotic behaviour}}
\begin{equation}
\phi(r)=\frac{2M\xi}{(\xi+1)r}+\cdots,
\end{equation}
{\bf{where exponentially decaying terms have been neglected. This shows that $e^{\phi(r)}\to 1$ as $r\to\infty$ and therefore, the manifold described by (\ref{guappofinal}) is Minkowski flat asymptotically away. We conclude this section by showing that our droplet allows for bound states of massive particles. By means of ($25.16$) in \cite{Fliessbach} we immediately find that the effective potential for the droplet is given by}}
\begin{equation}\label{effp}
U_{eff}(r)=\frac{e^{\phi(r)}}{2}\left(\epsilon+\frac{\ell^2}{r^2}\right), 
\end{equation}
{\bf{where $\epsilon$ and $\ell$ have been already defined in the previous section. Since $U_{eff}$ is nonnegative, a matching procedure with the effective potential for a Schwarzschild BH cannot be achieved in this case. However, if we insist that the total mass $M$ of the droplet coincides with the mass $M_{BH}$ of the black hole at the galactic centre and we introduce the rescalings $r^{*}=r/r_s$, $L=\ell/r_s$ and $H=h/r_s$ with $r_s=2M_{BH}$, we can rewrite $\phi$ and $B$ as follows}}
\begin{eqnarray}
\phi(r^{*})&=&\frac{1}{H^3}\int\frac{r^{*}e^{-\left(\frac{r^{*}}{H}\right)^{1/\xi}}}{B(r^{*})}\left[\frac{2}{\xi\Gamma(3\xi)}-3f(r^{*})\right]dr^{*},\\
B(r^{*})&=&1-\frac{{r^{*}}^2}{H^3}e^{-\left(\frac{r^{*}}{H}\right)^{1/\xi}}f(r^{*}),\quad
f(r^{*})=\sum_{k=0}^\infty\frac{(r^{*}/H)^{k/\xi}}{\Gamma(k+3\xi+1)},
\end{eqnarray}
{\bf{where we made use of $6.5.29$ in \cite{Abra}. From Fig.~\ref{mlaka} we see that in the regime of low $L$ the effective potential exhibits a minimum for which bound states for massive particles can form and a maximum corresponding to an unstable orbit. The situation is very different in the case of a massless particle where no bound states are allowed. We conclude with the observation that this model is not appropriate to reproduce the galactic motion of S-stars but it is nevertheless interesting because it indicates the possibility that a mass of dark matter modelled in terms of an anisotropic fluid with an Einasto energy density profile permits stable trajectories for massive particle over a large region as it can be seen in Fig.~\ref{mlaka}. Such a droplet albeit not suitable to model the central galactic object could possibly be formed in other parts of the galaxy. We conclude this section by estimating the value of the density and pressures at the centre of the droplet and comparing them with that for degenerate matter. More specifically, we consider the case of a neutron star with typical density $10^{17}$ Kg/m$^3$ and degenerate pressure of the order $10^{31}\div 10^{34}$ Pa while in our model we take $M=10M_\odot$. By means of (\ref{dpfE}) and (\ref{prrr}), we find that}}
\begin{equation}
\rho(0)=\frac{M}{4\pi h^3\xi\Gamma(3\xi)}\leq\frac{3.387 M}{4\pi h^3} ,\quad
p_r(0)=\frac{c^2}{3}\rho(0),
\end{equation}
{\bf{where in the expression for $\rho(0)$ we used the fact that the function $1/(\xi\Gamma(3\xi))$ has a global maximum for $\xi=0.1538$. We summarized the values of the density and radial pressure at the centre of the droplet in Table~\ref{presss}.}}
\begin{table}[!ht]
\caption{{\bf{Typical values of $\rho(0)$ and $p_r(0)$ at the centre of the droplet. The scaling factors $h_E$, $h_{B1}$ and $h_{B2}$ are chosen as in \cite{Einasto,deSalas}.}}}
\begin{center}
\begin{tabular}{ | l | l | l | l|l|}
\hline
$M=10M_\odot$                   &  $\rho(0)$ Kg/m$^3$      &  $p_r(0)$ Pa \\ \hline
$h_E=2.121\cdot 10^{-9}$ Kpc    &  $\leq 2.17\cdot 10^9$   &  $\leq 8.01\cdot 10^{23}$          \\ \hline
$h_{B1}=3.89\cdot 10^{-11}$ Kpc &  $\leq 4.36\cdot 10^{12}$& $\leq 1.30\cdot 10^{29}$          \\ \hline
$h_{B2}=1.43\cdot 10^{-5}$ Kpc  &  $\leq 8.85\cdot 10^{-5}$& $\leq 2.64\cdot 10^{12}$         \\ \hline
\end{tabular}
\label{presss}
\end{center}
\end{table}
\begin{figure}[!ht]\label{zabrezec}
\includegraphics[scale=0.35]{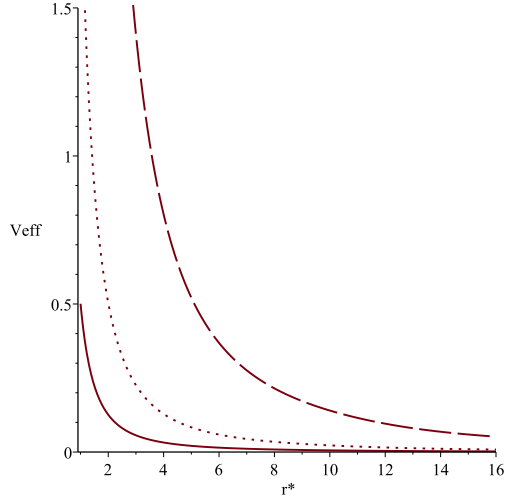}
\caption{\label{klmn}
  Plot of the effective potential (\ref{effp}) for $\epsilon=0$, $\xi=0.2$, $H=10$ with $L=1$ (solid line), $L=2$ (dotted line) and $L=5$ (long dashed line).}
\end{figure}

\section{Conclusions}\label{Sec6}
Motivated by the DM dominance of the galaxy we established a connection between DM and the BH physics by considering the Einasto density profile. We showed that starting with this profile and coupling it to an anisotropic energy-momentum tensor, it is possible to derive different black hole solutions by making certain choices for the underlying  EOS. In the case of an EOS of the form $p_r=-\rho$, we show that it is possible to construct a self-gravitating droplet or a BH depending on the values of the mass parameter $\mu$. If there is a horizon we could call such an object a fuzzy BH \cite{DavidePiero}. These objects made of DM are in nature different from the DM clumps constructed in \cite{Bosh} albeit the idea to connect the central galactic object to DM is similar.
We also discovered that the Einasto inspired black holes increases its Hawking temperature, as the horizon radius shrinks, until the temperature reaches a maximum after which the temperature decreases sharply and vanishes exactly at the radius of the extremal black hole. In both cases, a fuzzy droplet or a fuzzy BH, it is possible to obtain an effective potential which governs the equation of motion in such way  that the orbits will be as in the case of  a standard galactic BH. {\bf{If, instead of the previous EOS, we adopt a nonlocal EOS, it is possible to construct a self-gravitating droplet but it seems that a negative radial pressure cannot be  avoided. Moreover, an analysis of the effective potential shows the occurrence of bound states as well as the presence of an unstable orbit in the case of massive particle and low values of their angular momenta.}}

We also notice that the Einasto profile is a generalization of a Gaussian. Black Holes based on such density profile have been considered before. As a result, issues regarding the perturbation \cite{Pert} or exotic spacetime structure by considering a more general atlas \cite{SP} would proceed along similar lines.

{\bf{Finally, while finishing this manuscript, we found a paper \cite{Becerra} which is relevant to our work because it shares conclusions similar to those obtained here but by a different method. More precisely, \cite{Becerra} showed by a numerical simulation that if we replace the central supermassive BH by an object made of darkinos, this mass of DM would not only produce the same kinematics for S-stars but can also explain the G2 anomaly \cite{Park}.}}

\begin{acknowledgements}
We thank the anonymous referee for the critical reading and useful comments which helped improve the quality of the paper.
\end{acknowledgements}


\begin{thebibliography}{999}
\bibitem{BH} 
C. Bambi, {\it{Black Holes: A Laboratory for Testing Strong Gravity}}, Springer Nature Singapore (2017)
\bibitem{BHinterior1}
I. D. Novikov, {\it{The Internal Structure of Black Holes}} in ``Particle Physics and the Universe'', Proceedings of the 19th Adriatic Meeting, Sept. 2003, Springer Verlag Berlin Heidelberg (2005); P. R. Brady, {\it{The internal Structure of black holes}}, Prog. Theoret. Phys. Suppl. {\bf{136}}, 29 (1999); A. Bonnano, S. Droz, W. Israel and S. M. Morsink, {\it{Structure of the spherical black hole interior}},  Proc. Roy, Soc. A {\bf{450}}, 553 (1999); Y. Nomura, F. Sanches and S. J. Weinberg, {\it{The Black Hole Interior in Quantum Gravity}}, Phys. Rev. Lett. {\bf{114}}, 201301 (2015); H. Chakrabarty, A. Abdujabbarov, D. Malafarina and C. Bambi, {\it{A toy model for a baby universe inside a black hole}}, Eur. Phys. J. C {\bf{80}}, 373 (2020)
\bibitem{BHinterior2}
R. Brustein, A. J. M. Medved and K. Yagi, {\it{Discovering the interior of black holes}}, Phys. Rev. D {\bf{96}}, 124021 (2017)
\bibitem{profiles}
J. F. Navarro, C. S. Frenk, S. D. M. White and D. M. Simon, {\it{The Structure of cold dark matter halos}}, Astrophys. J. {\bf 462}, 563 (1996); A. Burkert, {\it{The structure of dark matter halos in dwarf galaxies}}, Astrophys. J. {\bf 447}, L85 (1995); H. Zhao, {\it{Analytical models for galactic nuclei}}, MNRAS {\bf 278}, 488 (1996)
\bibitem{nohair}
R. Ruffini and J. A. Wheeler, {\it{Introducing the Black Hole}}, Physics Today {\bf{24}}, 130 (1971); N. G\"urdebeck,  {\it{No-hair Theorem for Black Holes in Astrophysical Enviroments}}, Phys. Rev. Lett. {\bf{114}}, 151102 (2015); M. Heusler, {\it{Black Holes Uniqueness Theorems}}, Cambridge University Presss, Cambridge (2010) 
\bibitem{more}
 A. Borde, {\it{Regular Black Holes and Topology Change}}, Phys. Rev. D {\bf{55}}, 7615 (1999); C. Bambi. D. Malafarina and L. Modesto, {\it{Non singular quantum-inspired gravitational collapse}}, Phys. Rev. D {\bf{88}}, 044009 (2013); D. Malafarina and P. Joshi, {\it{Compact objects from gravitational collapse: an analytical toy model}}, Eur. Phys. J. C {\bf{75}}, 596 (2015)
\bibitem{Bosh}
K. Boshkayev and D. Malafarina, {\it{A model for a dark matter core at the Galactic Centre}}, MNRAS {\bf{484}}, 3325 (2019)
\bibitem{DMBH}
  X. Hernandez and W. H. Lee, {\it{An upper limit to the central density of dark matter haloes from consistency with the presence of massive central black holes}},  MNRAS {\bf{404}}, L61 (2010); M. I. Zeknikov and E. A. Vasiliev, {\it{Absorption of Dark Matter by a Supermassive Black Holes at the Galactic Centre: Role of the Boundary Conditions}}, Sov. Phys.  JETP {\bf{81}}, 85 (2005)
\bibitem{BHnoncomm}
P. Nicolini, {\it{Noncommutative Black Holes, The Final Appeal To Quantum Gravity: A Review}}, Int. J. Mod. Phys. A{\bf{24}}, 1229 (2009)
\bibitem{BHmodels2}  
 A. D. Sakharov, {\it{The initial stage of an expanding universe and the appearance of a nonuniform distribution of matter}}, Sov. Phys. JETP {\bf{22}}, 345 (1966); J. Bardeen, {\it{Non-singular general-relativistic gravitational collapse}} in Proceedings of ``The International Conference GR5'', Tiflis, USSR, 1996; I. G. Dymnikova, {\it{Vacuum non-singular black hole}}, Gen. Relat. Grav. {\bf{24}}, 235 (1992); I. G. Dymnikova, {\it{The cosmological term as a source of  mass}}, Class. Quant. Grav. {\bf{19}}, 725 (2002); I. G. Dymnikova, {\it{Spherical symmetric space-time with regular de Sitter center}},  Int. J. Mod. Phys. D {\bf{12}}, 1015 (2003); I. G. Dymnikova, {\it{Regular electrically charged vacuum structures with de Sitter centre in nonlinear electrodynamics coupled to general relativity}}, 
 Class. Quant.  Grav. {\bf{21}}, 4417 (2004); E. Ayon-Beato and A. Garcia, {\it{The Bardeen model as a nonlinear magnetic monopole}}, Phys. Lett. B {\bf{493}}, 149 (2000); J.Lemos and V. Zanchin, {\it{Regular black holes: Electrically charged solutions, Reissner-Nordstr\''om outside a de Sitter core}}, Phys. Rev. D {\bf 83}, 124005 (2011)
 \bibitem{Einasto}  
J. Einasto, {\it{On Galactic Descriptive Functions}},  Astron. Nachr. {\bf{291}}, 97 (1968); J. Einasto, {\it{The andromeda galaxy M 31: I. A preliminary model}}, Astrophysics {\bf{5}}, 67 (1969)
\bibitem{Navarro}
J. F. Navarro, E. Hayashi, C. Power et al., {\it{The inner structure of $\Lambda$ CDM haloes – III. Universality and asymptotic slopes}}, MNRAS {\bf{349}}, 1039 (2004)
\bibitem{Springel}
V. Springel, S. D. M. White, A. Jenkins et al., {\it{Simulations of the formation, evolution and clustering of galaxies and quasars}}, Nature {\bf{435}}, 629 (2005)
\bibitem{Mamon}
G. A. Mamon and E. L.  Łokas, {\it{Dark matter in elliptical galaxies – I. Is the total mass density profile of the NFW form or even steeper?}}, MNRAS {\bf{362}}, 95 (2005)
\bibitem{Cardone2005}
V. F. Cardone, E. Piedipalumbo and C. Tortora,  2005, {\it{Spherical galaxy models with power-law logarithmic slope}}, MNRAS {\bf{358}}, 1325 (2005)
\bibitem{Mer}
D. Merritt, A. W. Graham, B. Moore, J. Diemand, J. and B. Terzi$\acute{\mbox{c}}$, {\it{Empirical models for Dark Matter Halos. I. Nonparametric Construction of Density Profiles and Comparison with Parametric Models}}, AJ {\bf{132}}, 2685 (2006)
\bibitem{Ha}
E. Hayashi and S. D. M. White, {\it{Understanding the halo-mass and galaxy-mass cross-correlation functions}}, MNRAS {\it{388}}, 2 (2008)
\bibitem{Gao}
L. Gao, J. F. Navarro, S. Cole et al., {\it{The redshift dependence of the structure of massive $\Lambda$ cold dark matter haloes}}, MNRAS {\bf{387}}, 536 (2008)
\bibitem{Dhar}
B. K. Dhar and L. L.  Williams, {\it{Surface mass density of the Einasto family of dark matter haloes: are they Sersic-like?}}, MNRAS {\bf{405}}, 340  (2010)
\bibitem{Navarro2010}
J. F. Navarro, A. Ludlow, V. Springel et al., {\it{The diversity and similarity of simulated cold dark matter haloes}}, MNRAS {\bf{402}}, 21 (2010)
\bibitem{Chemin}
L. Chemin, W. J. G. de Blok and G. A. Mamon, {\it{Improved modeling of the mass distribution of disk galaxies by the Einasto Halo model}}, AJ {\bf{142}}, 109 (2011)
\bibitem{Retana}
E. Retana-Montenegro, E. Van Hese, G. Gentile, M. Baes and F. Frutos-Alfaro, {\it{Analytical properties of Einasto dark matter halos}}, A\&A {\bf{540}}, A70 (2012)
\bibitem{Ghez1}
A. M. Ghez et al., {\it{Stellar orbits around the Galactic Center Black Hole}}, ApJ {\bf{620}}, 744 (2005)
\bibitem{Ghez2}
A. M. Ghez et al., {\it{Measuring Distance and Properties of the Milky Way's Central Supermassive Black Hole with Stellar Orbits}}, ApJ {\bf{689}}, 1044 (2008)
\bibitem{Nobel}
It suffices here to give a web-site: https://www.nobelprize.org/prizes/physics/2020/press-release/
\bibitem{MazMo}
P. O. Mazur and E. Mottola, {\it{Gravitational vacuum condensate stars}}, PNAS {\bf{101}}, 9545 (2004)
\bibitem{ChiRe}
C. B. M. H. Chirenti and L. Rezzolla, {\it{How to tell a gravastar from a black hole}}, Class. Quantum Gravity {\bf{24}}, 4191 (2007)
\bibitem{RuBo}
R. Ruffini and S. Bonazzola, {\it{Systems of Self-Gravitating Particles in General Relativity and the Concept of an Equation of State}}, Phys.  Rev. {\bf{187}}, 1767 (1969)
\bibitem{ScMi}
F. E. Schunck and E. W. Mielke, {\it{General relativistic boson stars}}, Class. Quantum Gravity {\bf{20}}, R301 (2003)
\bibitem{Joshi}Tkachev
P.S. Joshi, D. Malafarina and R. Narayan, {\it{Equilibrium configurations from gravitational collapse}}, Class. Quantum Gravity {\bf{28}}, 235018 (2011)
\bibitem{BaMa}
C. Bambi and D. Malafarina, {\it{K$\alpha$ iron line profile from accretion disks around regular and singular exotic compact objects}}, Phys. Rev. D {\bf{88}}, 064022 (2013)
\bibitem{Chowd}
A. N. Chowdhury et al., {\it{Circular geodesics and accretion disks in the Janis-Newman-Winicour and gamma metric spacetimes}} Phys. Rev. D {\bf{85}}, 104031 (2012)
\bibitem{Kundt}
W. Kundt, {\it{ Galactic Nuclei}}, Astrophys. Space Sci. {\bf{235}}, 319 (1996); Erratum: ibidem {\bf{243}}, 263 (1997)
\bibitem{Ru1}
R. Ruffini, C. R. Arg\"{u}elles and J. A. Rueda, {\it{On the core-halo distribution of dark matter in galaxies}}, MNRAS {\bf{451}}, 622 (2015)
\bibitem{Ru2}
C. R. Arg\"{u}elles et al., {\it{Novel constraints on fermionic dark matter from galactic observables I: The Milky Way}}, Phys. Dark Universe {\bf{21}}, 82 (2018)
\bibitem{Sofue}
Y. Sofue, 2013, {\it{Rotation Curve and Mass Distribution in the Galactic Center —From Black Hole to Entire Galaxy}}, Publ. Astron. Soc. Jpn {\bf{65}}, 118 (2013)
\bibitem{Leu}
D. G. Levkov, A. G. Panin and I. I. Tkachev, {\it{Gravitational Bose-Einstein Condensation in the Kinetic Regime}}, Phys. Rev. Lett. {\bf{121}}, 151301 (2018)
\bibitem{Potter}
J. Stadel et al., {\it{Quantifying the heart of darkness with GHALO – a multibillion particle
simulation of a galactic halo}}, MNRAS {\bf{398}}, L21 (2009)
\bibitem{deSalas}
P. F. de Salas et al., {\it{On the estimation of the local dark matter density using the rotation curve of the Milky Way}}, J. Cosmol. Astropart. Phys. {\bf{10}}, 37 (2019)
\bibitem{Davies}
J. I. Davies, S. Phillipps, M. G. M. Cawson, M. J.  Disney E. J.  Kibblewhite, {\it{Low surface brightness galaxies in the Fornax cluster : automated galaxy surface photometry. III.}}, MNRAS {\bf{232}}, 239 (1988)
\bibitem{Caon}
N. Caon, M. Capaccioli and M. D’Onofrio, {\it{On the shape of the light profiles of early-type galaxies}}, MNRAS {\bf{265}}, 1013 (1993)
\bibitem{Don}
M. D’Onofrio, M. Capaccioli and N. Caon, {\it{On the shape of the light profiles of early-type galaxies-II. The $(D_n/A_e)-\langle\mu\rangle_e$ diagram}}, MNRAS {\bf{271}}, 523 (1994)
\bibitem{Cell}
S. A. Cellone, J. C. Forte and D. Geisler, {\it{A Morphological and Color Study of Fornax Low Surface Brightness Galaxies in the Washington System }}, ApJS {\bf{93}}, 397 (1994)
\bibitem{Andre}
Y. C. Andredakis, R. F. Peletier and M. Balcells, {\it{The shape of the luminosity profiles of bulges of spiral galaxies}}, MNRAS {\bf{275}}, 874 (1995)
\bibitem{Prug}
P. Prugniel and F. Simien, {\it{The fundamental plane of early-type galaxies: non-homology of the spatial structure}}, A\&A {\bf{321}}, 111 (1997)
\bibitem{Moll}
C. M\"{o}llenhoff and J. Heidt, {\it{Surface photometry of spiral galaxies in NIR: Structural parameters of disks and bulges }}, A\&A {\bf{368}}, 16 (2001)
\bibitem{Grah}
A. W. Graham and R. Guzm$\acute{\mbox{a}}$n, {\it{HST photometry of dwarf elliptical galaxies in Coma, and an explanation for the alleged structural  dichotomy between dwarf and bright elliptical galaxies}}, AJ {\bf{125}}, 2936 (2003)
\bibitem{Grah1}
A. W. Graham, D. Merritt, B. Moore, J. Diemand and B. Terzi$\acute{\mbox{c}}$, {\it{Empirical models for dark matter halos. II. Inner profile slopes, dynamical profiles, and $\rho/\sigma^3$}}, AJ {\bf{132}}, 2701 (2006)
\bibitem{Gad}
D. A. Gavotte, {\it{Structural properties of pseudo-bulges, classical bulges and elliptical galaxies: a Sloan Digital Sky Survey perspective}}, MNRAS {\bf{393}}, 1531 (2009)
\bibitem{gamma}
M. A. Chaudry and S. M. Zubair, {\it{On a Class of Incomplete Gamma Functions with Applications}}, Chapman \& Hall, Boca Raton, Florida (2002)
\bibitem{Prud}
A. P. Prudnikov, Yu. A. Brychkov and O. I. Marichev, {\it{Integral and Series}}, Vol. 2, Gordon and Breach Science Publishers (1986)
\bibitem{Abra}
M. Abramowitz and I. Stegun, {\it{Handbook of Mathematical Functions with Formulas, Graphs, and Mathematical Tables}}, Dover, New York, ninth Dover printing, tenth GPO printing edition (1964)
\bibitem{Piero}
  P. Nicolini, A. Smailagic and E. Spallucci, {\it{Noncommutative geometry inspired Schwarzschild black hole}}, Phys. Lett. B {\bf{632}}, 547 (2006)
\bibitem{DavidePiero}
  D. Batic and P. Nicolini, {\it{Fuzziness at the horizon}}, Phys. Lett. B {\bf{692}}, 32 (2010)
\bibitem{Mazur}
P. Mazur and E. Motolla, {\it{Surface tension and negative pressure interior of non-singular black holes}}, Class. Quant. Grav. {\bf{32}},  215024 (2015)
\bibitem{Dvali}
G. Dvali and C. Gomez, {\it{Black Hole Macro-Quantumness}}, arXiv:1212.0765
\bibitem{Dvali1}
G. Dvali and C. Gomez, {\it{Black Hole’s $1/N$ Hair}}, Phys. Lett. B {\bf{719}}, 419 (2013)
\bibitem{Dvali0}
G. Dvali and C. Gomez, {\it{Black Hole’s Quantum $N$-Portrait}}, Fortschr. Phys. {\bf{61}}, 742 (2013)
\bibitem{Dvali2}
G. Dvali and C. Gomez, {\it{Black Holes as Critical Point of Quantum Phase Transition}}, Eur. Phys. J. C {\bf{74}}, 2752 (2014)
\bibitem{Dvali3}
G. Dvali et al., {\it{Scrambling in the Black Hole Portrait}}, Phys. Rev. D {\bf{88}}, 124041, (2013)
\bibitem{Dvali4}
G. Dvali and C. Gomez, {\it{Landau-Ginzburg limit of black hole’s quantum portrait: Self-similarity and critical exponent}}, Phys.  Lett. B {\bf{716}}, 240 (2012)
\bibitem{Dvali5}
G. Dvali and C. Gomez, {\it{Quantum Compositeness of Gravity: Black Holes, AdS and Inflation}}, J. Cosmol. Astropart. Phys. {\bf{1401}}, 023 (2014)
\bibitem{Giddings}
S. B. Giddings, {\it{Hawking radiation, the Stefan Boltzmann law, and unitarization}}, Phys. Lett. B {\bf{754}}, 39 (2016)
\bibitem{Liberati}
  R. Dey, S. Liberati and D. Pranzetti, {\it{The black hole quantum atmosphere}}, Phys. Lett. B {\bf{774}}, 308 (2018)
\bibitem{Felten}
J. E. Felten and R. Isaacman, {\it{Scale factor $R(t)$ and critical values of
cosmological constant $\Lambda$ in Friedmann universes}}, Rev. Mod. Phys.  {\bf{34}}, 689 (1986)
 \bibitem{negativeP}
A.R. Imre, H. J. Maris and P.R. Williams (eds.), {\it{Liquids under negative pressure}}, NATO science series, Springer Science + Buisness Media, Dodrecht (2002)
\bibitem{Fliessbach}
T. Fliessbach, {\it{Allegemeine Relativitätstheorie}}, (Elsevier, New York, 2006)
\bibitem{Urry}
C. M. Urry and P. Padovani, {\it{Unified Schemes for Radio-Loud Active Galactic Nuclei}}, Publ. Astron. Soc. Pac. {\bf{107}}, 803 (1995)
\bibitem{ABN}
I. Arraut, D. Batic and M. Nowakowski, {\it{A non commutative model for a mini black hole}}, Class. Quantum Gravity {\bf{26}}, 245006 (2009)
\bibitem{Olver}
R. A. Askey and R. Roy, R., {\it{Series Expansions}}, in  F. W. J. Olver et al., {\it{NIST Handbook of Mathematical Functions}}, Cambridge University Press, Cambridge (2010)
\bibitem{Hern1}
H. Hernandez, L. A. Nunez and U. Picric, {\it{Nonlocal Equation of State in General Relativistic Radiating Spheres}}, Class. Quant. Grav. {\bf{16}}, 871 (1999)
\bibitem{Hern2}
H. Hernandez and L.A. Nunez, {\it{Nonlocal Equation of State in Anisotropic Static Fluid Spheres in General Relativity}},  Can. J. Phys. {\bf{82}}, 29 (2004)
\bibitem{Ab1}
H. Abreu, H. Hernandez and L.A. Nunez, {\it{Sound Speeds, Cracking and Stability of Self-Gravitating Anisotropic Compact Objects}}, Class. Quant. Grav. {\bf{24}}, 4631 (2007)
\bibitem{PIEROBOSS}
P. Nicolini and E. Spallucci, {\it{Noncommutative geometry inspired wormholes and dirty black holes}}, Class. Quant. Grav. {\bf{27}}, 015010 (2010)
\bibitem{Pert}  
D. Batic, N.G. Kelkar, M. Nowakowski and K. Redway, 
{\it{Perturbing microscopic black holes inspired by noncommutativity}}, Eur.Phys. J. C {\bf{79}}, 581 (2019)
\bibitem{SP}
I. Arraut, D. Batic and M. Nowakowski,  {\it{Maximal Extension of the Schwarzschild Spacetime Inspired by Noncommutative Geometry}},  J. Math. Phys. {{\bf 51}}, 022503 (2010)  
\bibitem{Becerra}
E. A. Becerra-Vergara et al., {\it{Hinting a dark matter nature of Sgr A$^{*}$ via the S-stars}}, MNRAS: Letters, slab051 (2021)
\bibitem{Park}
J. H. Park et al., {\it{No asymmetric outflows from Sagittarius A$^*$ during the pericenter passage of the gas cloud G2}}, A\&A {\bf{576}}, L16 (2015)
\end{thebibliography}
\end{document}